\documentclass[12pt]{article}
\usepackage{graphicx}
\usepackage{epsfig}
\usepackage{latexsym}
\usepackage{latexsym}
\usepackage{amssymb}
\usepackage{amsmath}
\newcommand{\labell}[1]{\label{#1}}
\newcommand{\reef}[1]{(\ref{#1})}

\DeclareSymbolFont{AMSb}{U}{msb}{m}{n}
\DeclareMathSymbol{\IN}{\mathbin}{AMSb}{"4E}
\DeclareMathSymbol{\IZ}{\mathbin}{AMSb}{"5A}
\DeclareMathSymbol{\IR}{\mathbin}{AMSb}{"52}
\DeclareMathSymbol{\Q}{\mathbin}{AMSb}{"51}
\DeclareMathSymbol{\II}{\mathbin}{AMSb}{"49}
\DeclareMathSymbol{\IC}{\mathbin}{AMSb}{"43}
\DeclareMathSymbol{\IP}{\mathbin}{AMSb}{"50}
\DeclareMathSymbol{\IH}{\mathbin}{AMSb}{"48}
\DeclareMathSymbol\IA{\mathalpha}{AMSb}{"41}
\DeclareMathSymbol\IS{\mathalpha}{AMSb}{"53}

\def\Q{{\cal Q}}

\begin{document}

\begin{flushright}
{\tt arXiv:0707.4303}
\end{flushright}
{\flushleft\vskip-1.35cm\vbox{\psfig{figure=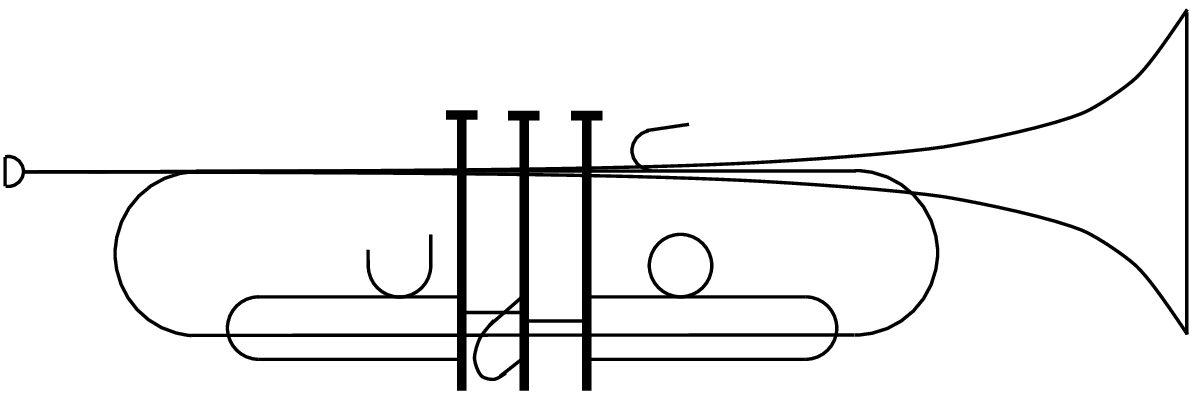,height=0.45in}}}
\bigskip
\bigskip
\bigskip
\bigskip
\bigskip

\begin{center} {\Large \bf Heterotic Coset Models}

\bigskip

{\Large\bf of}

\bigskip

{\Large\bf   Microscopic Strings and Black Holes}

\end{center}

\bigskip \bigskip \bigskip

\centerline{\bf Clifford V. Johnson}

\bigskip
\bigskip

  \centerline{\it Department of Physics and Astronomy }
\centerline{\it University of
Southern California}
\centerline{\it Los Angeles, CA 90089-0484, U.S.A.}

\bigskip

\centerline{\small \tt  ${}^1$johnson1 [at] usc.edu}

\bigskip
\bigskip


\begin{abstract} 
\noindent Following a recent conjecture by Lapan, Simons and Strominger, we
revisit and discuss an intrinsically heterotic class of conformal
field theories, emphasizing their Lagrangian construction as
asymmetrically gauged WZW models, which may be useful in several
applications to the study of supersymmetric strings and black holes in
heterotic and type~II string theory compactified on $T^6$ and
$K3\times T^2$ respectively. In these cases, the leading supergravity
geometry is singular, but higher order corrections remove this
singularity in a way that is consistent with, for example the
non--zero entropy for the black holes that these strings form after
wrapping on an additional circle. The conformal field theories have
the right structure to capture the features of the supergravity
analysis, and possess precisely the microscopic target spaces
required. We describe in detail the model with AdS$_3\times S^2$
geometry, which is conjectured by Lapan {\it et.  al.} to represent a
fundamental heterotic string in five dimensions, and then propose
conformal field theories which are potential candidates for the
microscopic geometry of heterotic strings in $D$ dimensions, with
target space AdS$_3\times S^{D-3}$. We also discuss some conformal
field theories that give microscopic AdS target spaces in various
dimensions.

\end{abstract}
\newpage \baselineskip=18pt \setcounter{footnote}{0}

\section{Introduction}
\label{sec:introduction}
In recent times, supersymmetric black hole solutions of ${\cal N}=2$
and ${\cal N}=4$ supergravity in four dimensions have been of
considerable interest. The latter, the initial focus of this paper,
can be studied as D4--D0 bound states in type~II string theory
compactified on $K3\times T^2$ (the D4--brane world--volume wraps the
$K3$), or in heterotic string theory compactified on $T^6$, where the
black holes arise as fundamental heterotic
strings\cite{Susskind:1993ws,Russo:1994ev,Sen:1995in} wrapped on one
of the circles of the compactification with the inclusion of
Kaluza--Klein momentum.  These two--charge black holes, while
classically having horizons of zero area (and in fact are naively
singular at their core), turn out to have much more structure when
studied beyond the leading order in the small $\alpha^\prime$
expansion. The corrections remove the singularity and yield a smooth
AdS$_2\times S^2$ geometry at the core\cite{Dabholkar:2004dq}. The
microstate counting (leading to the entropy, including the leading
contribution and a family of corrections) for these black holes on the
heterotic string side is remarkably successful\cite{Dabholkar:2004yr},
and suggests (as was the case for other successful microstate
studies\cite{Strominger:1996sh,Strominger:1997eq}) that the conformal
field theory whose properties control the correspondence between the
heterotic counting and the black hole geometry has an holographic dual
spacetime.

A very natural question to ask (as emphasized by Strominger recently
in a talk at Strings 2007\cite{Strominger:Strings}) is whether there
is an accessible description of the geometry of the fundamental
heterotic string which itself becomes the black hole.  This is
motivated by the analogy with the D1--D5--momentum system which is
used for the successful microstate counting of the three--charge
five--dimensional black holes in type~II string
theory\cite{Strominger:1996sh,Strominger:1997eq} --- there is a
spacetime at the core which is asymptotically AdS$_3\times S^3$, which
is holographically dual to the $(1+1)$--dimensional theory (derived
from the D1--D5 world--volume) doing the state counting.  The nature
of this new spacetime is particularly interesting, since while the
leading order geometry of the fundamental string (the source that
plays the role of the D--branes in the situation in hand) has a null
singularity at the core, quite tantalizingly the string coupling is
weak there. As a reminder, the form of the geometry for $N$ strings
lying along the direction $x_1$ in $D$ non--compact directions is:
\begin{eqnarray}
  \label{eq:stringmetric}
  &&ds^2_h=H^{-1}(-dt^2+dx_1^2)+dr^2+r^2d\Omega^2_{D-3}+ds^2_{T^{10-D}}\ ;\nonumber\\
&&e^{-2(\Phi-\Phi_0)}=H\ ;\qquad B_{01}=-H^{-1}\ ;\nonumber\\
&&H=1+N\left(\frac{r_h}{r}\right)^{D-4}\ ; \qquad r_h^{D-4}=\frac{r_1^6}{V_{10-D}}\ ;\qquad r_1^6=T\frac{16\pi G_{10}}{6\Omega_7}\ , 
\end{eqnarray} 
where $r$ is the radial coordinate in $D$ dimensions,
$d\Omega^2_{D-3}$ is the metric on the round unit $S^{D-3}$, and
$r_1^6$ is the standard constant that ensures that the solution has
$N$ units of fundamental heterotic string tension
$T=(2\pi\alpha^\prime)^{-1}$. Here, $G_{10}$ is Newton's constant in
ten dimensions, $\Omega_7$ is the volume of a unit seven--sphere, and
$V_{10-D}$ is the volume of the torus, $T^{10-D}$, on which we've
compactified. From this it is clear that as we move to the singular
core at $r=0$, the string coupling goes as:
\begin{equation}
  \label{eq:stringcoupling}
  g_s\sim \frac{1}{N^\frac12}\left(\frac{r}{r_h}\right)^{\frac{D-4}{2}}\ ,
\end{equation}
which is generically quite small (and can be tuned as small as one
desires by sending $N$ large, although this is not necessary in this
example). The curvature is clearly blowing up in this near--core
limit, and so one should expect $\alpha^\prime$ corrections to the
solution to be important, and can be expected to modify the story.

The $\alpha^\prime$ corrections have been shown to reveal an
AdS$_3\times S^2$ corrected geometry\cite{Castro:2007sd}, for the
unwrapped $T^5$ case of the infinite straight fundamental heterotic
string. We need to go beyond supergravity to the full heterotic string
theory to fully study this situation. There is additional hope for
success since the heterotic string theory, our arena of study, has
only NS--NS sources, which are readily accessible in conformal field
theory, in contrast to the R--R sources of type~II.

So in view of these encouraging signs it is prudent to seek a
tractable conformal field theory representing this microscopic
heterotic geometry, capturing the physics of the string theory in this
regime\footnote{While this manuscript reporting our results was being
  completed, we learned of a paper\cite{Dabholkar:2007gp} by Dabholkar
  and Murthy which presents results in this area which may be
  related.}. This conformal field theory is logically distinct from
the conformal field theory on the world--volume of the stretched or
wrapped heterotic strings that become the black hole. The latter is
the one that would be the holographic dual of the microscopic
spacetime at the core. However, there is evidently a trilogy of
dictionaries allowing translations between any two of the three
systems. One is the standard world--sheet/space--time correspondence
of string theory, while another would be of the now familiar AdS/CFT
holographic type. The third one which is implied is a new
correspondence that does not seem to involve gravity.

For the simpler case of the unwrapped string, a recent conjecture of
Lapan, Simons, and Strominger (announced by the latter at Strings
2007\cite{Strominger:Strings}) suggested\footnote{Since an earlier
  version of this manuscript appeared, a paper by Lapan, Simons, and
  Strominger has now been submitted to the arXiv\cite{Lapan:2007jx}.}
that the (apparently) difficult $S^2$ part of the conformal field
theory is a special uncharged case of the asymmetric orbifold
presented\cite{Giddings:1993wn} by Giddings, Polchinski and Strominger
(GPS) in 1993, in the context of four dimensional non--supersymmetric
magnetically charged black holes in heterotic string theory. It turns
out that there is a non--trivial conformal field theory for the
angular sector even when the charge of those black holes vanishes, and
this is the special case that Lapan {\it et. al.}  conjecture can then
be tensored with an $SL(2,\IR)$ conformal field theory in order to
represent the straight fundamental heterotic string in the $T^5$
compactification.

This conjecture is compelling, and while several aspects of the
supergravity aspects of this study are still being considered by Lapan
et. al., (but see footnote~2) we wish to firmly discuss the conformal
field theory aspects further in this paper, and confirm that the
suggestion makes sense.  Furthermore, the idea leads us naturally to
present several more conformal field theories and make some natural
conjectures and suggestions concerning them.

The first mission of this paper is to emphasize that the full $S^2$
conformal field theory (and others in its class) is in fact quite easy
to define in a full path integral formalism (it is arguably more
natural to present it this way than as an orbifold) which may well be
useful for future computations in this context. We present this in
section 2, and this is entirely a review of the ``heterotic coset
model'' construction presented\cite{Johnson:1994jw} by the author in
1994, the prototype of which was shown to be the GPS model.  Next, in
section~3, insights gained from the method immediately lead us to new
conformal field theories that we conjecture represent fundamental
heterotic strings in heterotic string theory compactified on
$T^{10-D}$ to $D$ dimensions, where the ``microscopic'' geometry (its
radii are again of order $\alpha^\prime$) is AdS$_3\times S^{D-3}$.
The cosets in these cases are non--Abelian. In section~4 we discuss
briefly the analogous constructions for conformal field theories with
AdS$_{p+2}$ target spaces, where $p\geq0$. Again their radius is
frozen to be of order~$\alpha^\prime$. It is not clear what their role
is in the current context, but since they may turn out to be relevant,
we present them here.  Generalizations to cases with fewer
supersymmetries, where the non--angular geometry has several higher
order corrections in $\alpha^\prime$, and a non--trivial radial
dependence for the dilaton, are also discussed in the concluding
section~5.

\section{The GPS Model from a Lagrangian Perspective}
As originally suggested in ref.\cite{Johnson:1994jw}, the GPS monopole
model\cite{Giddings:1993wn} model is very naturally defined as a
gauged WZW model\cite{Witten:1983ar,Gawedzki:1988hq,Gawedzki:1988nj}
with a number of (relatively) unusual features that give it an
intrinsically heterotic characteristic. The construction works as
follows (we will write relatively few formulae and instead emphasize
the concepts since the original paper\cite{Johnson:1994jw} is quite
explicit). Starting with an $SU(2)$ WZW with coordinates $(z,{\hat
  z})$, and field $g(z,{\bar z})\in SU(2)$, which has an
$SU(2)_L\times SU(2)_R$ global symmetry, gauge a purely right--acting
$U(1)$ subgroup, $g\to gh$, where $h\in U(1)_R$.  The resulting model
is classically anomalous, which is to say that a gauge invariant
Lagrangian cannot be written for the model.  Nevertheless, it is
fruitful to exploit the freedom\cite{Witten:1991mm} to introduce a two
dimensional gauge field with components $A_z,A_{\bar z}$, and couple
it to $g$ in such a way that under a $U(1)_R$ gauge transformation,
the Lagrangian changes by an amount proportional to:
\begin{equation}
  \label{eq:anomaly}
  \delta I=\frac{1}{8\pi}\int d^2\!z\, F_{z{\bar z}}\ .
\end{equation}
In fact, all two dimensional gauge anomalies that we will consider,
classical or quantum (as will arise from fermions), can be written in
this way, which is key. In the conventions we will choose here, the
constant of proportionality for the gauging of the $U(1)_R$ is
precisely $k$, the level of the WZW.  This is because we choose the
$U(1)_P$ to be generated by $i\sigma_3/2$, where $\sigma_3$ is the
standard (real, diagonal) third Pauli matrix, and the constant is,
more generally, $k{\rm Tr}[\sigma_3^2]/2$. (For a more general
discussion, see ref.\cite{Johnson:1994kv}.)

To complete the model, we must introduce fermions. As this is the
heterotic string, the left movers must be coupled in a way that gives
us left--moving supersymmetry. (Note that we have exchanged left and
right here relative to the choices made in ref.\cite{Johnson:1994jw}.
This will give us a positive value for $k$ at the special point we are
interested in.) There are two such fermions, and they are naturally
defined as taking their values on the coset $SU(2)/U(1)$. This fixes
very specifically the coupling of those fermions, determining their
charge under the $U(1)_R$. They are of course anomalous under the
$U(1)_R$, and in our conventions the anomaly is simply $-2$, a
contribution of $-1$ for each fermion. We may couple in some number of
right--moving fermions, with charge~$Q$ under the $U(1)_R$, where we
are much more free to choose the value of $Q$ since we are not
constrained by the requirement to get a right--moving world--sheet
supersymmetry. Picking the most natural quantity of fermions, two,
their anomaly is simply $2Q^2$, where there is a contribution of $Q^2$
for each fermion, and the opposite sign is due to the opposite
chirality.

The complete $(0,2)$ model then has the three sectors. We can cancel
the quantum and classical anomalies against each other if we satisfy
the equation:
\begin{equation}
k=2(1-Q^2)= -2Q_+Q_-\ ,
\labell{eq:anomalycondition}
\end{equation}
where $Q_\pm=Q\pm1$. Here $k$ will not go negative since we will be
choosing $Q$ to vanish shortly. For $Q>1$, one can change the sign by
simply exchanging the left and right moving fermions, or by acting on
the left with the gauge action. To make things simple, we can
henceforth write $k=2|Q_+Q_-|$ in general formulae.

The complete model is now written as a Lagrangian definition, with
$g$, the fermions from the left and the right, and the gauge field
$A_z, A_{\hat z}$ all coupled together. It is gauge invariant, and
defines a consistent conformal field theory. While it is a $(2,0)$
model on the world--sheet (as is guaranteed because this is an
$SU(2)/U(1)$ coset, which is Kahler\cite{Kazama:1988qp}), it is not
spacetime supersymmetric in general.  Modular invariance requires $Q$
to be integer in our units (matching the fact that it is a $U(1)$
monopole in spacetime), but this is inconsistent with the world--sheet
condition on charges that promotes $(2,0)$ supersymmetry to spacetime
supersymmetry. For the black hole application that GPS had in mind,
(and the various generalizations to a host of interesting spacetimes
in refs.\cite{Johnson:1994ek,Johnson:1995ga}), this model is combined
with a radial direction $\sigma$ and a time direction $t$ defining a
non--trivial ``radial sector'' $(\sigma,t)$ conformal field theory,
also arising from a gauged WZW (usually based on $SL(2,\IR)/U(1)$, as
in refs.\cite{Bars:1989ph,Witten:1991yr}). The latter sector has a
level $k^\prime$ which is linearly related to $k$ by the overall
condition on the central charge of the total model.  The spacetime
geometry of the black hole can be reliably read off (carefully --- see
ref.\cite{Johnson:1994jw} for subtleties involving fermionic
back--reaction arising from cancelling a classical anomaly against a
quantum one) the resulting heterotic sigma model in the large $k$
(small $\alpha^\prime$) limit, or equivalently (because of
equation~\ref{eq:anomalycondition}), when the charge $Q$ is large.

We are not interested in spacetime solutions that have $U(1)$ gauge
(as opposed to Kaluza--Klein) monopole charge $Q$ in spacetime,
however. That $U(1)$ is a subgroup of the full heterotic string gauge
group.  This is not of interest to us here, and so we should in fact
set $Q=0$. Interestingly, there is a non--trivial solution (dubbed the
``remnant'', by GPS) to the equation, $k=2$. This system is therefore
not appropriate for describing a solution which has a large (as
compared to $\alpha^\prime$) geometrical footprint in spacetime. It is
a geometry that is microscopic, from the supergravity perspective.
Happily (looking at just this angular sector on its own for now), the
model still has the geometry of an $S^2$. There are no $\alpha^\prime$
corrections, as can be seen in various ways, the most straightforward
among them being: (1) the whole model can be written (by fixing a
gauge and bosonizing the fermions into a single extra bosonic field)
as an asymmetrically acting $\IZ_2$--orbifold of an $SU(2)$ WZW --- the
original presentation of GPS for $Q=0$, and (2) the model is spacetime
supersymmetric when $Q=0$.

To be slightly more explicit, the  round $S^2$ that results from the
construction has the familiar metric:
\begin{equation}
  \label{eq:stwometric}
  ds^2=k(d\theta^2+\sin^2\theta d\phi^2)\ ,
\end{equation} written in terms of the the standard angles
$\theta$ and $\phi$, which originate as part of the set of Euler
angles $(\theta,\phi,\psi)$ of the $S^3=SU(2)$, where a group element
can be written:
\begin{equation}
  \label{eq:groupelement}
  g=e^{i\phi\sigma_3/2}e^{i\theta\sigma_2/2}e^{i\psi\sigma_3/2}\ ,\qquad 0\leq\theta\leq\pi\ ,\,\,0\leq\phi\leq2\pi\ ,\,\, 0\leq\psi\leq4\pi\ .
\end{equation}
The angle $\psi$ is fibred over the $S^2$ in the standard Hopf manner
to make an $S^3$. The right $U(1)$ action we discussed earlier is
entirely on $\psi$, and after writing the gauged model, a natural
gauge in which to work while studying the physics is $\psi=\mp\phi$,
(where the ``$\mp$'' choice refers to the Northern or Southern
hemispheres of the $S^2$) which leaves the two scalar fields
$X^1=\phi, X^2=\theta$, together with the left and right moving
fermions, which are equivalent after bosonization to another scalar we
can call $X^3$. The action is, after integrating out the world--sheet
gauge fields (a procedure which is exact since there is no field
dependence in the $A_zA_{\bar z}$ term):
\begin{eqnarray}
  \label{eq:fullmodel}
  &&I=\frac{k}{4\pi}\int\!d^2z\Biggl\{G^{S^2}_{\mu\nu}\partial_z X^{\mu}\partial_{\bar z}X^{\nu} +\frac{1}{Q_+^2}\left(\partial_zX^3-2Q_+A^M_\mu\partial_zX^\mu\right)\left(\partial_{\bar z}X^3-2Q_+A^M_\mu\partial_{\bar z}X^\mu\right)\nonumber\\&&\hskip2.6cm-\frac{2A^M_\mu}{Q_+}\left(\partial_{\bar z}X^3\partial_z X^\mu-\partial_z X^3 \partial_{\bar z}X^\mu\right)\Biggr\}\ ,
\end{eqnarray}
where $G_{\mu\nu}^{S^2}$ is the metric on the unit round $S^2$,
$k=2|Q_+Q_-|$\, and the {\it spacetime} background $ A^M_\mu$ has only
one non--zero component given by $2A^M_\phi=\pm1-\cos\theta$, where
the ``$\pm$'' choice refers to the Northern or Southern hemispheres of
the $S^2$. In a fermionic presentation of the content of $X^3$, the
left--movers couple covariantly to $A^M_\mu$ with charge unity and the
right--movers with charge $Q$, the latter being the gauge monopole
charge that we will set to zero for our purposes, as already stated. A
quick way to see that this, fermions and all, is also equivalent (as
shown by GPS) to a {\it bosonic} $SU(2)$ WZW (up to a discrete
identification to match the periodicity of the fields to the $2\pi$ of
$X^3$) is to write $X^3=Q_+(\psi\pm\phi)$, from whence some algebra
will return one to the standard WZW action with $S^3$ in the metric
and a torsion term induced by the $S^3$ volume form
$H\sim\sin\theta\,d\theta\wedge d\phi\wedge d\psi$ which can be
locally written as $H=dB\sim(\pm1-\cos\theta)\, d\phi\wedge d\psi$.

The power and clarity obtained from recasting the GPS model in this
way as a heterotic coset with a Lagrangian definition (with the
explicit formulae for the world--sheet gauge couplings given in
ref.\cite{Johnson:1994jw}) should not be underestimated. It allows for
many more models to be easily and quite intuitively defined by simply
picking subgroups to gauge as dictated by one's geometrical
requirements (for example, the freedom to leave an entire $SU(2)_L$
untouched in the prototype GPS model guarantees the rotational
invariance of the whole model, and hence a round $S^2$ metric) and
then coupling in fermions in a manner which allows one to cancel the
total anomaly. Furthermore, various quantities in the conformal field
theory are readily extracted (see e.g. ref.\cite{Berglund:1995dv}, for
these types of model), sometimes more easily than other methods, by
using the path integral definition of the conformal field theory
afforded by the method. It is to be expected that this may well be
likely useful in the current black hole application.

Returning to the GPS monopole theory, the suggestion of Lapan {\it et.
  al.}  that it should be tensored with an $SL(2,\IR)$ conformal field
theory as a candidate for the conformal field theory of the
microscopic AdS$_3\times S^2$ geometry of the (corrected) core of the
stretched heterotic string in five dimensions makes perfect sense.
There are some pleasant consequences to be derived when this is all
put together: The level of $SL(2,\IR)$, $k^\prime$ is again fixed by
the total condition on the central charge. We add three fermions (in
Lie($SL(2,\IR)$)) on the left for supersymmetry, and the trivial $T^5$
sector. The central charge in each sector is of the form\footnote{In
  the first version of this manuscript, we incorrectly
  shifted\cite{Knizhnik:1984nr} the level $k$ by ${\rm g}$ in the
  formula following, which led to an inconsistency. We thank Atish
  Dabholkar and Sameer Murthy for a question about this issue.}:
\begin{equation}
c=\frac{k{\rm dim G}}{k+{\rm g}}\ ,
  \label{eq:centralcharge}
\end{equation}
where ${\rm g}$ is the Coxeter number of the group (which is 2 here
for each factor).  As is by now
familiar\cite{Bars:1989ph,Witten:1991yr}, there is a continuation
$k^\prime\to-k^\prime$ for $SL(2,\IR)$ to achieve a $(-++)$ signature.

The condition on the left that $c_L=15$ yields the
same equation that the condition on the right that $c_R=26$, which is:
\begin{equation}
\frac{3k^\prime}{k^\prime-2}+\frac{3k}{k+2}=6\ ,
  \labell{eq:kcondition}
\end{equation}
which is solved by $k^\prime=k+4$. Since $k=2$, we must have
$k^\prime=6$. The AdS$_3$ geometry of the angular sector is again
microscopic, and uncorrected due to being a group manifold.

Finally, it is worth noting that since the microscopic squared radius
of the $S^2$ part of the microscopic string is $k=2$, we have the
result that the area of the resulting four dimensional black hole
(obtained by wrapping the string on an additional circle) is $8\pi$ in
dimensionless units, in string frame. To get to Einstein frame, we
need the value of the four dimensional dilaton, which contains the
information about~$N$, the number of wrapped strings. According to
ref.\cite{Dabholkar:2004dq}, the attractor
equations\cite{Ferrara:1996dd} give $e^{-2\Phi}=N^{\frac12}$, and so
in Einstein frame, we get the area to be $A=8\pi\sqrt{N}$, which after
dividing by two (not four in this case\cite{Dabholkar:2004dq}) is
indeed the quantum corrected result for the entropy, as can be
computed in the holographically dual conformal field theory on the
world--sheet of the $N$ wrapped strings (with one unit of
Kaluza--Klein momentum) or in the supergravity using the Wald entropy
formula\cite{Dabholkar:2004yr,Wald:1993nt,Jacobson:1993vj}.  That this
conformal field theory gives precisely the right value for the radius,
and leaves room for no other, is encouraging.

\section{Fundamental Heterotic Strings in Other Dimensions}
It would be neglectful to stop at this point, since there is a very
natural set of heterotic cosets to explore which may be of relevance
to stretched heterotic string sources.  The point is that in $D$
dimensions, the relevant sphere surrounding a string is $S^{D-3}$. It
is again very easy to see how to design the conformal field theory
having an exact microscopic geometry of such spheres, using the same
heterotic coset method, and the nice fact that:
\begin{equation}
  \label{eq:cosetspheres}
  \frac{SO(n)}{SO(n-1)}\sim S^{n-1}\ .
\end{equation}
Simply gauge according to this coset, with a pure a right action, and
the model will be guaranteed to have the required $SO(n)$ global
symmetry arising from the untouched $SO(n)_L$ symmetry of the
conformal field theory\footnote{The case $n=4$ is interesting since
  then the target is $S^3$. One might have imagined that simply using
  an $SU(2)$ WZW would have sufficed for this example, but there would
  be no reason to have its size be of order $\alpha^\prime$ since no
  anomaly condition would arise to restrict $k$.}.

Our example of section~2 was the case $n=3$, for which the gauge group
is Abelian. Here, the gauging is non--Abelian, which is a significant
difference from the lower dimensional case. As before, however, the
classically anomalous gauging must be compensated for by the quantum
anomaly of the left--moving fermions one would add for supersymmetry.
This will result in a specific value for the level~$k$ as we shall
discuss further shortly. This is all then to be combined with the
radial sector.  For the radial sector, one would just use the
$SL(2,\IR)$ conformal field theory at level $k^\prime$ again, for the
stretched string's presumed AdS$_3$ microscopic geometry.

Again, we can check the central charge, which will give us a condition
linking $k^\prime$ and $k$. For $c_L=15$ or $c_R=26$ (with $n=D-2$):
\begin{equation}
\frac{3k^\prime}{k^\prime-2}+\frac{n(n-1)}{2}\frac{k}{k+{\rm g}}-\frac{(n-1)(n-2)}{2}+\frac12\times3+\frac12\times(n-1)=\frac32\times(n+2)\ ,
  \labell{eq:kconditionbig}
\end{equation}
where ${\rm g}=n-2$ or $n-1$ (the Coxeter number of the group $SO(n)$
for $n$ even or odd, respectively) and so we get the condition, after
some algebra:
\begin{equation}
k^\prime=2+ \frac{12(k+{\rm g})}{n(n-1){\rm g}}\ .
  \label{eq:highercentralcharge}
\end{equation}
The consistency condition from the anomaly equation from asymmetric
gauging to give the sphere gives $k={\cal A}_n$, where ${\cal A}_n$ is
a pure number which depends on $n$ in a manner which results from the
details of the non--Abelian embedding of the action of the gauge group
and how the left--moving supersymmetry fermions in the coset couple.
There are ${\rm dim} (S^{n-1})=n-1$ such fermions, and so we expect
that their anomaly (which has coefficient $-{\cal A}_n$) will not grow
any faster than linearly in $n$. (It is tempting to simply write
${\cal A}_n=n-1$, assuming a contribution of $-1$ for each fermion in
the normalization we've been using, but this should be proven.)

So finally, we can define a consistent model (at least as far as
gauging and total central charge is concerned) which it is natural to
suggest supplies a conformal field theory definition of the stretched
heterotic string in $D$ dimensions by setting $k={\cal A}_n$ and
\begin{equation}
k^\prime=2+ \frac{12({\cal A}_n+{\rm g})}{n(n-1){\rm g}}\ .
  \label{eq:highercentralcharge2}
\end{equation}
Our model has the expected microscopic target space AdS$_3\times
S^{D-3}$, since the radii are both of order~$\alpha^\prime$.  The
dilation will again be as small as desired, with no coordinate
dependence since the quadratic terms in the gauge fields for the
non--Abelian gauge action defining the conformal field theory can be
written in a way that does not involve any of the other
fields\cite{Witten:1991mm}. The Jacobian which induces the dilaton
coupling in the sigma--model will therefore be field (and hence
spacetime coordinate) independent. For the same reason, we expect that
the target space geometry is exact.

\section{Higher Dimensional Anti--de Sitter Geometries}
Given what we did in the previous sections, it is not hard to see how
to design conformal field theories which have microscopic AdS$_{p+2}$
geometries, for $p\geq0$. It follows quite straightforwardly since
anti--de Sitter spacetimes are easily obtained by analytically
continuing spheres, and thus the coset of
equation~\reef{eq:cosetspheres}, appropriately continued, will serve
to define the desired spacetimes ($p=n-3$).  It is not clear what the
role of such target spaces is, (nor what the detailed spectrum of the
conformal field theories thus defined is), but the construction is
quite natural.

The case $p=0$, AdS$_2$, is actually a special case of one of several
models proposed in ref.\cite{Johnson:1994jw}, and of a model
independently obtained from the GPS--type orbifold perspective in
ref.\cite{Lowe:1994gt}. We gauge a spacelike $U(1)_R$ subgroup of the
$SL(2,\IR)$ WZW (with right--moving fermions) in a manner analogous to
the procedure we carried out for $SU(2)$.

The construction works most simply for this example by analytically
continuing much of the discussion in section~2 of the $S^2$ conformal
field theory.  By analytically continuing $SU(2)$ to $SL(2,\IR)$, or
otherwise, it is easy to see that our $U(1)$ generators are again
naturally written in terms of Pauli matrices (divided by two), and so
we will choose $\sigma_3/2$ as our generator.  An example of a
continuation is:
\begin{equation}
  \label{eq:continuation}
  \theta\to i\sigma\ ,\,\, \phi\to i t\ ,\,\, k^\prime\to -k^\prime\ ,
\end{equation}
yielding in the heterotic sigma model the AdS$_2$ metric:
\begin{equation}
  \label{eq:adsmetric}
  ds^2=k^\prime(d\sigma^2-\sinh^2\sigma dt^2)\ . 
\end{equation}
Other continuations will yield metrics on AdS$_2$ that cover different
coordinate patches as desired. 

In the same natural normalization as we used before (where gauge
quantities and fermions are naturally living in the Lie algebra of the
group, or in the case of cosets, the difference between the Lie
algebras of the group and the subgroup), the classical anomaly due to
gauging is now $k^\prime$, while the quantum anomaly for the
supersymmetry fermions (there are two, naturally living in the coset,
which is two dimensional) is again $-2$.  Therefore, we can again
construct a consistent model with $k^\prime=2$.  By analogy with the
$S^2$ case, it is easy to see that the theory (after fixing a gauge)
can be rewritten as a bosonic $SL(2,\IR)$ WZW (up to an
identification), for which there are no corrections to the metric.

\section{Conclusion}
This construction of the heterotic conformal field theories with
target AdS$_3\times S^2$ is a very natural description of the
microscopic heterotic string, as suggested by Lapan {\it et.
  al.}\cite{Strominger:Strings}.  By formulating it here as a
heterotic coset model, the generalizations of section~3 to the
important cases of stretched strings in higher dimensions are quite
straightforward. As we have noted, for the higher dimensional models
the coset construction is non--Abelian.

There are other models in ref.\cite{Johnson:1994jw} which can be
revisited in the light of this proposal as models of (generically
non--supersymmetric) microscopic heterotic strings. There are several
two dimensional $SL(2,\IR)$ models with gauge actions on both the left
and right which were still asymmetric. The anomaly from this can again
be cancelled against the anomaly from the supersymmetry fermions, and
the whole model tensored with the $Q=0$ GPS monopole representing the
$S^2$ theory. This would give a model with non--trivial
$\alpha^\prime$ corrections, and a dilaton that varies in the
$\sigma$--direction. This might be a model of a wrapped heterotic
string in four dimensions which becomes a non--supersymmetric black
hole, and as such might make contact with some of the discussions of
refs.\cite{Giveon:2004zz,Giveon:2005mi,Giveon:2005jv,Giveon:2006pr}.

Many other interesting models present themselves for reconsideration,
such as those that arose from the twisting together of the angular and
radial sectors\cite{Johnson:1994jw}, resulting in a non--trivial
stringy Taub--NUT solution (it came with dyonic charges and
non--trivial NUT-- and $H$--charge from $G_{t\phi}$ and $B_{t\phi}$
components). A search for a consistent microscopic solution of the
various anomaly equations that ensured the consistency of that
solution seems to yield interesting solutions that deserve further
study.

\section*{Acknowledgments}
CVJ wishes to thank Nick Halmagyi for conversations and for bringing
Strominger's Strings 2007 talk to his attention. He would also like to
thank Sergei Gukov and Dave Morrison each for a conversation.  CVJ
also thanks Atish Dabholkar and Sameer Murthy for a comment on an
earlier version of the manuscript of this paper. This work is
supported by the DOE. This work was done at the Aspen Center for
Physics. Thanks to the staff at the Center for an excellent working
environment.

\providecommand{\href}[2]{#2}\begingroup\raggedright\endgroup

\end{document}